\documentclass[conference]{IEEEtran}
\IEEEoverridecommandlockouts
\usepackage{cite}
\usepackage{amsmath,amssymb,amsfonts}
\usepackage{algorithmic}
\usepackage{graphicx}
\usepackage{textcomp}
\usepackage{multirow}
\usepackage[normalem]{ulem}
\usepackage{soul}
\usepackage{xcolor}
\usepackage[none]{hyphenat}
\usepackage{tabularx}

\setstcolor{red}

\def\BibTeX{{\rm B\kern-.05em{\sc i\kern-.025em b}\kern-.08em
    T\kern-.1667em\lower.7ex\hbox{E}\kern-.125emX}}
\begin{document}

\title{RAG-PRISM: A Personalized, Rapid, and Immersive Skill Mastery Framework with Adaptive Retrieval-Augmented Tutoring}

\author{
\IEEEauthorblockN{
Gaurangi Raul\IEEEauthorrefmark{1},
Yu-Zheng Lin\IEEEauthorrefmark{2},
Karan Patel\IEEEauthorrefmark{2},
Bono Po-Jen Shih\IEEEauthorrefmark{5},
Matthew W. Redondo\IEEEauthorrefmark{2}\\
Banafsheh Saber Latibari\IEEEauthorrefmark{2},
Jesus Pacheco\IEEEauthorrefmark{4},
Soheil Salehi\IEEEauthorrefmark{2},
Pratik Satam\IEEEauthorrefmark{2}\IEEEauthorrefmark{3}
}

\IEEEauthorblockA{
\IEEEauthorrefmark{1}College of Information Science, University of Arizona, Tucson, AZ, USA\\
\IEEEauthorrefmark{2}Department of Electrical and Computer Engineering, University of Arizona, Tucson, AZ, USA\\
\IEEEauthorrefmark{3}Department of Systems and Industrial Engineering, University of Arizona, Tucson, AZ, USA\\
\IEEEauthorrefmark{4}Department of Industrial Engineering, University of Sonora, Hermosillo, Mexico\\
\IEEEauthorrefmark{5}Leonhard Center for Enhancement of Engineering Education, The Pennsylvania State University, University Park, PA, USA\\
}
\IEEEauthorblockA{
Emails: \{graul\IEEEauthorrefmark{1}, yuzhenglin\IEEEauthorrefmark{2}, karanpatel\IEEEauthorrefmark{2}, mredondo245\IEEEauthorrefmark{2}, banafsheh\IEEEauthorrefmark{2}, ssalehi\IEEEauthorrefmark{2}, pratiksatam\IEEEauthorrefmark{2}\IEEEauthorrefmark{3}\}@arizona.edu;\\
jesus.pacheco@unison.mx\IEEEauthorrefmark{4}; bps5848@psu.edu\IEEEauthorrefmark{5}
}
}

\maketitle

\begin{abstract}
 The rapid digital transformation of Fourth Industrial Revolution (4IR) systems is transforming today's workforce needs, increasing skill set gaps, especially for the older workforce. With an increasing emphasis on STEM skill sets like robotics, automation, artificial intelligence (AI), and security, the workforce will have to be re-skilled and up-skilled to meet future industry needs. While re-skilling/up-skilling a massive workforce, these programs have to be mindful of trainee's diverse backgrounds, pedagogy styles, and motivations to increase student persistence, retention, and success; to ensure rapid and cost-effective workforce development while ensuring skill building through experiential learning. \\
To address these challenges, we present an adaptive tutoring framework that explores the usage of generative artificial intelligence (AI) combined with Retrieval-Augmented Generation (RAG)s to generate personalized training for each students learning needs. Our framework uses a combination of document hit rate and Mean Reciprocal Rank (MMR) to personalize and optimize the training for the trainee. The framework's personalization is evaluated against a human-generated training to evaluate the framework's quality of personalization through source content alignment, and relevance metrics. We apply the proposed framework for 4IR cybersecurity learning, through the creation of a synthetic question-answer (QA) dataset emulating trainee behavior, while the RAGs are optimized on a curated cybersecurity learning materials corpus. The proposed framework, is evaluated for its new training generation, by comparison with a set of manually curated queries to represent realistic student interactions. The framework's responses are generated through multiple large language models (LLMs) including \texttt{GPT-3.5} and \texttt{GPT-4} variants, which are evaluated for content alignment, and relevance (faithfulness), with \texttt{GPT-4} having the best performance with a relevancy score of 87\%, and 100\% content alignment. Thus, our preliminary evaluation shows that this dual-mode approach allows the adaptive tutor to serve as both a new personalized topic recommender, providing a novel approach to provide rapid, personalized learning for 4IR learning and workforce development needs.
\end{abstract}


\section{Introduction}
The rise of Industry 4.0 (I4.0) has brought about complex cyber-physical systems (CPS) deeply integrated with artificial intelligence (AI), automation, and ubiquitous connectivity\cite{philbeck2018fourth}. The growing intricacy and connectivity of these I4.0 systems has significantly increased the complexity of these modern smart manufacturing systems, exposing new and evolving skillset gaps especially of STEM topics. Once such skillset gap is the need for cybersecurity training, where the increased complexity of such I4.0 systems, has exposed new attack surfaces, making them increasingly vulnerable to cyberattacks. As cyber threats evolve, it becomes essential to rapidly and cost-effectively train a skilled cybersecurity workforce capable of addressing these challenges. However, current cybersecurity training methods, particularly online programs, are not equipped to meet the unique demands of 4IR systems, which require specialized knowledge and hands-on training experiential learning \cite{satam2023cps,zachos2024vehicle,lin2025personalized,ghimire2024interactive}. Furthermore, the scale of training programs needed to up-skill and re-skill large, diverse groups of people from different areas, backgrounds, and learning needs, adding another layer of complexity \cite{carnevale2018balancing}.
Traditional cybersecurity education often struggles to accommodate learners from various backgrounds, each with differing levels of expertise \cite{lin2024prism}. This mismatch results in knowledge gaps that hinder effective learning and training. Although scalable online education programs are beneficial for broadening access, they fall short when it comes to providing personalized, adaptive learning experiences that can cater to the specific needs of individuals\cite{vykopal2022smart}, especially when faced with the constantly changing landscape of cyber threats \cite{satam2020wids,pacheco2020artificial,ghimire2025hwrex,satam2015anomaly,ghimire2025hardware}. 

 Furthermore, the current labor shortages will require new cybersecurity I4.0 workforce creation programs or reskilling/upskilling programs to have higher student retention and persistence while meeting the programs' learning objectives, a challenge difficult to overcome \cite{schwab2024fourth}. Researchers have shown that student background and upbringing play a role in student persistence through STEM programs, wherein data shows that it is especially challenging for students from marginalized communities like underrepresented minorities (URMs) to succeed. These difficulties can be attributed to a lack of access to high-quality education throughout the student’s formative years (pre/middle/high schools), creating a cyclic set of knowledge dependencies that are difficult to meet \cite{ogbu2022voluntary,dong2020understanding, wladis2015stem}. 

To address these challenges, this research explores an adaptive cybersecurity tutor that using generative artificial intelligence (AI) combined with Retrieval-Augmented Generation (RAG)s generates personalized training for each students learning needs. The proposed Adaptive tutor uses a combination of document hit rate and Mean Reciprocal Rank (MMR) to personalize and optimize the training for the trainee, with domain specific content generated via the RAG, optimized for the trainee's learning needs through the use of generative AI. The framework's personalization is evaluated against a human-generated training to evaluate the framework's quality of personalization through source content alignment and relevance metrics.

In this context, these are the unique contributions
of our work:

\begin{itemize}
    \item \textbf{Personalized Content Adaptation:} We present a framework that dynamically personalizes learning materials based on each student's background and domain-specific needs, leveraging sentiment-informed LLM interactions within a VR-based Digital Twin learning environment.

    \item \textbf{Contextualized Instruction Delivery:} The system tailors learning trajectories by aligning queries and responses to each learner’s specific topic of interest, enhancing contextual relevance and engagement.

    \item \textbf{Integrated RAG-Based Training Pipeline:} The framework integrates LlamaIndex-based Retrieval-Augmented Generation (RAG) and OpenAI’s GPT models to deliver adaptive cybersecurity training content.

    \item \textbf{Quantitative Evaluation of Retrieval and Generation:} To assess the effectiveness of the proposed RAG-based LLM learning system, we explore the following two research questions: \\
    1) In response to learner queries, to what extent can the system accurately retrieve the relevant course materials, as measured by hit rate and mean reciprocal rank (MRR), and faithfully generate relevant answers, as measured by faithfulness and relevancy? \\
    2) What is the comparative performance of different LLMs when integrated into the system, based on faithfulness, relevancy, and response length? \\
    Our evaluation shows that GPT-4 achieves a faithfulness score of 1.00 and a relevancy score of 0.93, outperforming other models. Retrieval effectiveness is demonstrated with a hit rate and MRR both reaching 1.00 in representative queries.
\end{itemize}

The rest of the paper is organized as follows. In Section \ref{sec:related_work}, reviews related work on the use of LLMs in education. Section \ref{sec:Architecture_framework} introduces the integration of sentiment analysis with digital twin observations from VR learning interface, followed by a sentiment-aware, RAG-enhanced LLM learning system and the overall architecture design. Section \ref{sec:Methodology} presents our implementation using LlamaIndex, hybrid question generation, and experimental results based on faithfulness and relevancy evaluation metrics. Finally, Section \ref{sec:Conclusion} concludes the study.

\section{Related work} \label{sec:related_work}

\subsection{LLMs in Education}
Large Language Models (LLMs), such as GPT and Llama, have demonstrated transformative potential in natural language understanding and generation. Unlike traditional rule-based systems or early-generation NLP chatbots that rely heavily on handcrafted features and rigid pattern matching \cite{lin2018development,ayanouz2020smart}, LLMs exhibit emergent capabilities including zero-shot reasoning, contextual adaptation, and fluent dialogue generation. These attributes enable LLMs to function as general-purpose tutors, dynamically responding to diverse learner inputs without predefined scripting.

For example, the PRISM (Personalized, Rapid, and Immersive Skill Mastery) framework proposed by Lin et al. \cite{lin2024prism} combines LLM-based zero-shot sentiment analysis with immersive virtual reality (VR) modules to provide a highly interactive and emotionally responsive learning environment. The sentiment analysis pipeline quantifies the engagement of the learner by transforming qualitative teacher-student dialogues into structured numerical indicators, enabling the system to adjust instructional strategies in response to emotional and cognitive states. Experimental results demonstrate that LLMs maintain strong zero-shot classification performance even in challenging informal settings involving internet slang, thereby ensuring robust affective inference across diverse communication styles. In addition to this, the virtual reality (VR) modules in the PRISM framework serve as an interactive learning interface that supports experiential education. These modules incorporate finite automata to implement adaptive difficulty progression, allowing the learning environment to respond dynamically to student performance. Furthermore, the VR-based Digital Twin models are constructed using photogrammetry techniques \cite{alhamadah2024photogrammetry}, which preserve geometric and visual fidelity while enabling realistic simulation of industrial systems. Together, these components allow the system to deliver a personalized, emotionally responsive, and task-aligned learning experience that integrates both affective sensing and hands-on skill development.

Despite these advancements, the initial implementation of GraphRAG within PRISM exhibits limitations in both scalability and quantitative evaluation. Although GraphRAG improves domain alignment by integrating structured knowledge retrieval, its static graph-based architecture lacks the flexibility required for seamless expansion to new domains and topics. Moreover, there remains a gap in systematically evaluating the outputs of such systems, particularly in terms of user specific, faithfulness\cite{jacovi2020towards} to source materials, and relevance to learner queries. Addressing these shortcomings requires more modular retrieval mechanisms and robust evaluation pipelines that can ensure both adaptive personalization and reliable performance metrics.

Our work bridges this gap by proposing a framework that not only integrates sentiment analysis, but also an LLM-driven learning environment with evaluation metrics, faithfulness and relevancy to assess the quality of LLM-generated responses. This approach ensures that personalized learning experiences contextual accuracy and real-time adaptability.

\subsection{Retrieval-Augmented Generation (RAG)}
Retrieval-Augmented Generation (RAG) is a hybrid natural language processing approach that enhances language model responses by retrieving relevant external documents during inference, combining dense document retrieval with generative models to produce contextually grounded and accurate outputs \cite{lewis2020retrieval}. Lewis et al. \cite{lewis2020retrieval} showed that RAG-style models significantly improve performance on knowledge-intensive tasks by fusing retrieval with generation in an end-to-end pipeline. As given, the powerful abilities of RAG in providing the latest and helpful auxiliary information, Retrieval-Augmented Large Language Models (RA-LLMs) have emerged to harness external and authoritative knowledge bases, rather than solely relying on the model's internal knowledge, to augment the quality of the generated content of LLMs \cite{fan2024survey}. It effectively combines the parameterized knowledge of LLMs with non-parameterized external knowledge bases, making it one of the most important methods for implementing LLMs \cite{gao2023retrieval}. Offloading the demand for accuracy and specificity of domain-specific knowledge to external sources reduces the sensitivity of generated outputs to the parameters and training quality of LLMs \cite{wan2025RAG}. This makes it more viable to use smaller, more efficient models that use not only fewer parameters but also lower-precision data formats, such as 8-bit integers instead of traditional 32-bit floats. In scenarios where data privacy is of utmost importance, improved efficiency in computing resources and energy consumption enables the on-premises deployment of RAG-enhanced LLM systems by small organizations \cite{ieva2024RAG}.

\begin{figure}[!h]
  \centering    \includegraphics[width=\columnwidth]{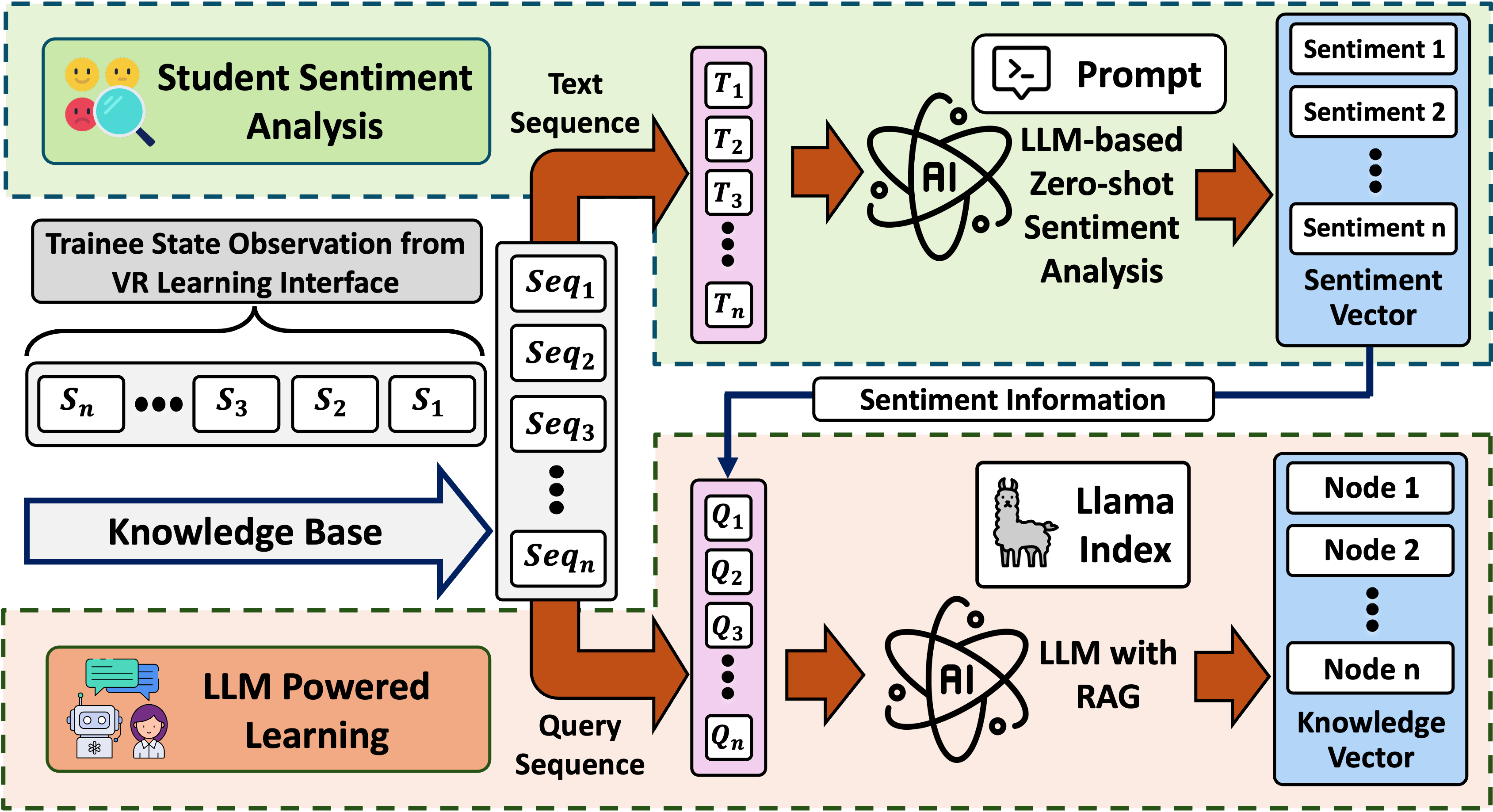}
    \caption{RAG-PRISM Framework. This figure illustrates a system that integrates LLM-based sentiment analysis with RAG-enhanced knowledge retrieval to support adaptive and personalized education.}
    \label{fig:Architecture framework}
\end{figure}

\section{RAG-PRISM Framework} \label{sec:Architecture_framework}
Building upon the foundational PRISM framework, we propose RAG-PRISM: A Personalized, Rapid, and Immersive Skill Mastery Framework with Adaptive Retrieval-Augmented Tutoring, as illustrated in Figure~\ref{fig:Architecture framework}. The proposed RAG-PRISM framework extends the original PRISM architecture \cite{lin2024prism} by incorporating retrieval-augmented generation (RAG) to improve the personalization of instructional content in response to learner sentiment and contextual input. This enhanced framework integrates core components necessary for providing sentiment-aware and context-sensitive tutoring, effectively overcoming the rigidity of traditional digital learning systems through real-time adaptation to both emotional and cognitive learning states. In this study, we focus on the retrieval and generation layer, implementing AI-powered instructional support guided by RAG. To ensure instructional quality, we adopt evaluation metrics such as faithfulness and relevancy, thus establishing a foundation for more effective and responsive AI-driven education.

\subsection{System Design}
This system represents a hybrid pipeline that integrates student sentiment analysis, LLM powered guidance, and knowledge retrieval. It is designed to enable personalized experiential learning within virtual reality-based Digital Twin environments.

\subsubsection{ Student Sentiment Analysis Module}
The top half of the workflow begins with the VR-based Digital Twin (DT) system monitoring the student’s learning states to effectively support the student's conducting experiments in that environment, through student behavior analysis by analyzing their interactions with the DTs, and the student sentiments, by analyzing their audio and text interactions with AI based instructors. Therefore, the PRISM tracks the student's learning through the following:
\begin{itemize}

\item \textbf{Input - Trainee State Observation from VR Learning Interface:}
Captures the dialogue content and state observations (e.g., learning activities and task completion patterns) from the VR learning interface. 

\item \textbf{Textual Sequence Extraction:}
These observations are converted into text sequences ($T_1$, $T_2$, ..., $T_n$), which represent student interaction logs or responses. It serves as an input for the multistage pipeline.

\item \textbf{LLM-based Zero-Shot Sentiment Analysis:}
These sequences are passed through a LLM (e.g., GPT-4), prompted using zero-shot sentiment analysis techniques. This eliminates the need for extensive training data and makes the system adaptable across domains. Here, it interprets the emotional undertones of the student's responses and
generates a sentiment vector\cite{lin2025personalized}.

\item \textbf{Output - Sentiment Vector:}
The model generates a sentiment vector capturing the emotional tone, confidence, and engagement level for each interaction. This vector is critical for tailoring the learning experience which is then passed to the second stage.
\end{itemize}

\subsubsection{LLM-Powered Personalized Learning Module}
This stage represents the adaptive learning side based on sentiment feedback.

\begin{itemize}
    \item \textbf{Query Sequence + Sentiment Vector:}
Based on the knowledge base (i.e., structured PDFs, educational content) it is converted into Queries. The query sequence consists of queries and information which is passed from the sentiment vector to it. 

\item \textbf{LlamaIndex + RAG (Retrieval Augmented Generation):}
These queries are processed using LlamaIndex, which connects LLMs with an external knowledge base. This enables the model to retrieve domain-relevant content before generating a response, leading to higher accuracy and contextual alignment.

\item \textbf{Knowledge Vector:}
The retrieved content is embedded into a knowledge vector, which represents highly relevant educational content personalized to the learner’s context and emotion.
\end{itemize}

Personalized learning plays a vital role in the success of this workflow, especially when scaling education across diverse learners. RAG-PRISM by incorporating real-time sentiment analysis, remains attuned to a learner’s emotional and cognitive state, helping maintain motivation and focus. This adaptive approach functions much like a skilled human tutor, dynamically adjusting content based on both knowledge gaps and emotional signals, which in turn enhances retention and comprehension. Moreover, the model fosters greater educational equity by offering high-quality, responsive instruction to students from underrepresented or underserved backgrounds without requiring access to expensive lab resources or constant human intervention.

\begin{figure}[!t]
  \centering    \includegraphics[width=\columnwidth]{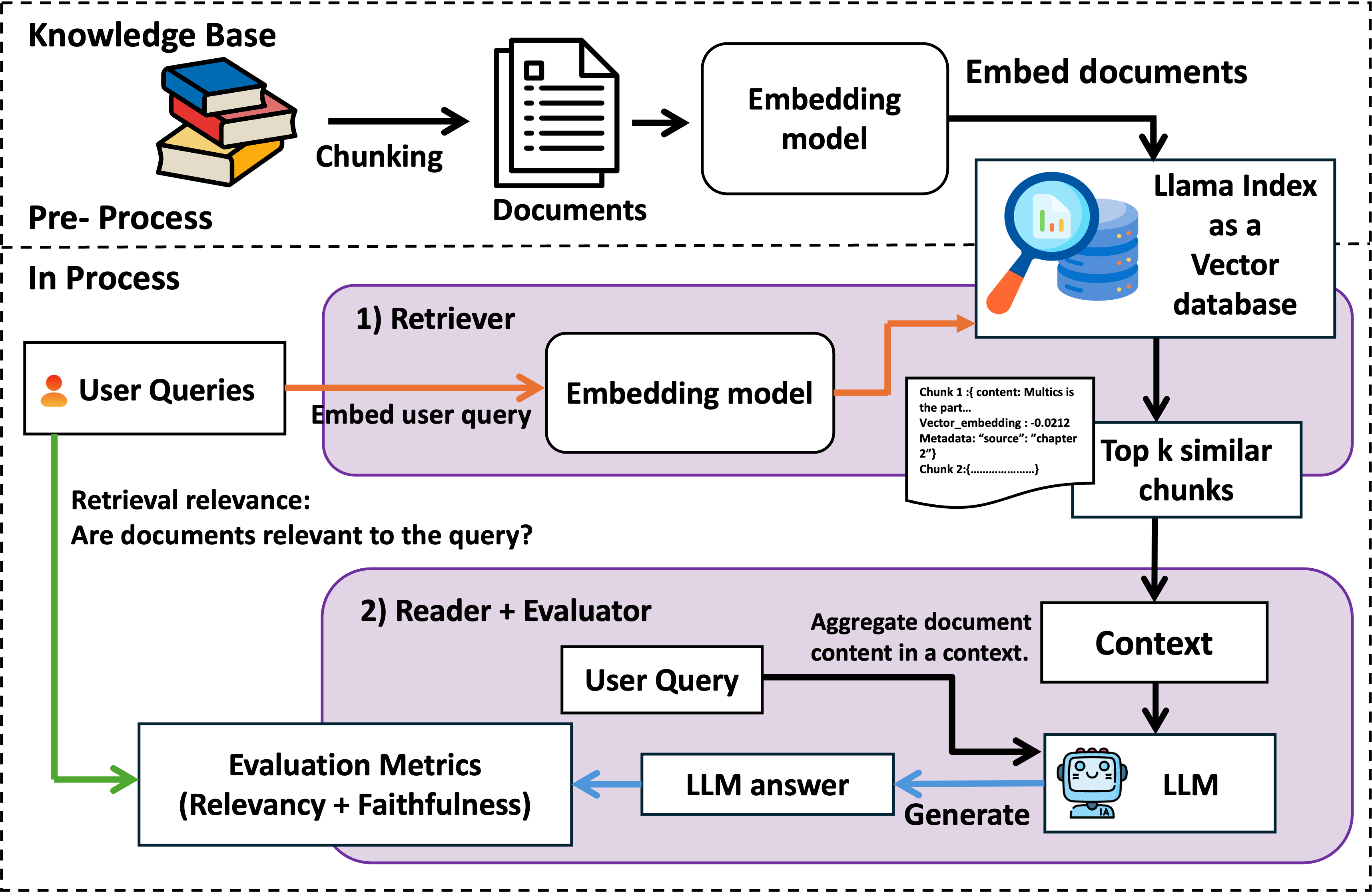}
    \caption{A LlamaIndex-Based Pipeline for Query Answering and Evaluation}
    \label{fig:Framework}
\end{figure}

\section{Experiment Methodology} \label{sec:Methodology}
\subsection{Experimental Setup}
To evaluate the proposed system, we detail the architecture and implementation pipeline, including the tools, libraries, and platforms used. The implementation spans several modular components: document ingestion, hybrid question generation, RAG, and evaluation.

 \subsubsection{\textbf{LLama Indexing Serving}}
LlamaIndex a flexible and modular data framework that bridges unstructured/structued data sources with LLMs for efficient retrieval-based applications. LlamaIndex simplifies the development of RAG pipelines by handling document ingestion, node parsing, semantic search indexing, and prompt construction, all through a high-level API. 

Some use cases include the following:
\begin{itemize}
    \item Question-Answering Chatbots (commonly referred to as RAG systems, which stands for "Retrieval-Augmented Generation")
    \item Document Understanding and Extraction
    \item Autonomous Agents that can perform research and take actions. 
\end{itemize}

Why llama Index is used and what does it reinforce?
\begin{itemize}
    \item It is managing the entire retrieval-to-prompt-building process.

    \item Abstracts away low-level logic, like chunking, filtering, and formatting.
    \item It acts as the middleware between the vector store and OpenAI’s models, simplifying RAG implementation.
\end{itemize}

\subsubsection{Hybrid Question Generation Module}
To evaluate and enhance the accessibility of the indexed material, as illustrated in Figure \ref{fig:Evaluation Framework}, we incorporated a hybrid QA generation approach with two parallel pathways (i.e., synthetic and manual) that  not only surfaces key concepts but also contributes to real-time knowledge enhancement for the LLM:

\begin{itemize}
    \item Synthetic QA generation: Leveraging document-ingested chunks, the model uses LLMs to auto-generate question-answer (QA) pairs that simulate learner-level curiosity and comprehension checks. These pairs are fed back into the query engine and evaluated using metrics like Mean Reciprocal Rank (MRR) and Document Hit Rate.
    This loop enables analysis of how well the retriever identifies relevant content and how grounded the model’s outputs are in the database, helping to ensure that generated answers remain both accurate and contextually aligned with the knowledge base.
    \item Manual Queries insertion: Curated queries are added to reflect the kinds of things real students might ask, including tricky or less obvious ones. These inputs serve as a control mechanism helping benchmark the system's retrieval accuracy and generative robustness, so the model stays useful and relevant to the actual course content.
\end{itemize}

This dual strategy of synthetic and manual queries generation, aids in extending the model's retrieval-guided reasoning and better aligning system outputs with pedagogical goals.
Studies, have explored LLM-centered educational tutoring frameworks where QA generation directly enhances learning experiences by tailoring output to a learner’s comprehension gaps. This notion of retrieval-grounded content creation is especially important when leveraging LLMs for adaptive instruction

\subsubsection{Llama Index based-RAG Pipeline}
A typical RAG pipeline is illustrated in Figure \ref{fig:Framework}. Unlike closed-book LLMs, which rely solely on their internal parameters, RAG models access external sources of truth to improve factual grounding and contextual relevance. Such as given GPT's reliance on pretraining data, it initially lacks the capacity to provide updates on recent developments. RAG bridges this information
gap by sourcing and incorporating knowledge from external databases.
The RAG system comprises two primary components: Retrieval and Generation. The retrieval component aims to extract relevant information from various external knowledge sources. Indexing organizes documents to facilitate efficient retrieval, using either inverted indexes for sparse retrieval or dense vector encoding for dense retrieval \cite{khattab2020colbertefficienteffectivepassage}  \cite{douze2024faiss} \cite{wei2022emergentabilitieslargelanguage} .The searching component utilizes these indexes to fetch relevant documents on the user’s query, often incorporating the optional re-rankers \cite{chen2024benchmarking}  to refine the ranking of the retrieved documents. 

In this case, it creates a database from relevant documents collected from the knowledge base (i.e., structured PDF on a contextual topic)
to the user’s query. These documents, combined with the original queries, form a comprehensive prompt that empowers LLMs to generate a well-informed answer.

\begin{figure}[!h]
  \centering    \includegraphics[width=1\linewidth]{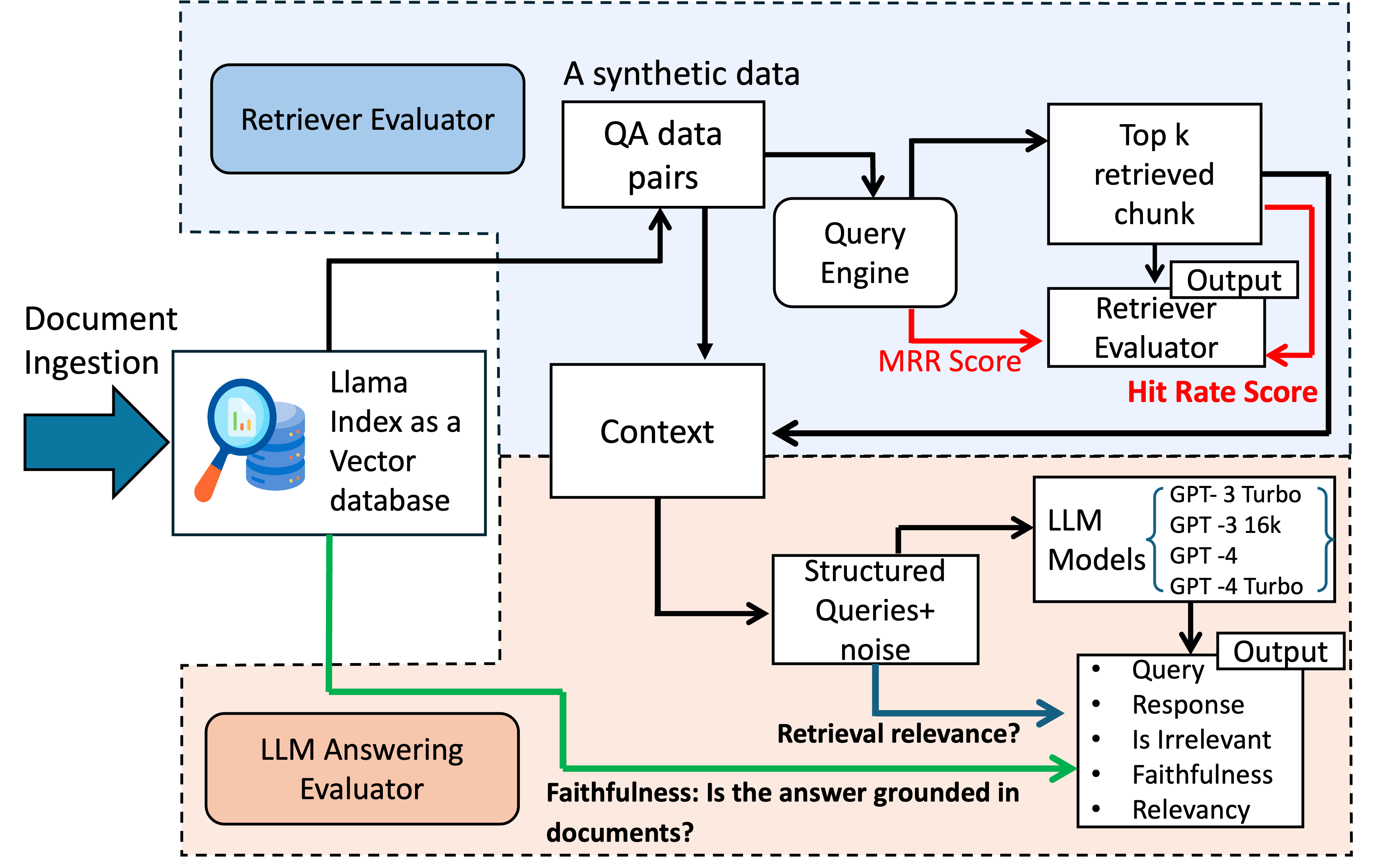}
    \caption{Hybrid Evaluation Framework}
    \label{fig:Evaluation Framework}
\end{figure}

The RAG system operates through a set of fundamental phases that structure the end-to-end retrieval and generation process:
\begin{itemize}
    \item \textbf{Step 1: Acquisition} - Raw educational content (e.g., structured PDFs) is extracted using tools and optionally expanded to API sources. LlamaIndex document loaders handle the integration of this content.

    \item \textbf{Step 2: Indexing} - The preprocessed text is fragmented and converted into vector embeddings via OpenAI models. LlamaIndex uses indexing structures (i.e., VectorStoreIndex) to create searchable representations.

    \item \textbf{Step 3: Storage} - Indexes and metadata are stored to maintain context and prevent recomputation. LlamaIndex supports storing these in-memory.

    \item \textbf{Step 4: Querying} - During inference, user queries are embedded and routed for semantic retrieval process. LlamaIndex query engine handles top-k retrieval, this enables retrieved context to be packaged into prompts and passed to LLMs for grounded response generation.

    \item \textbf{Step 5: Evaluation} - To monitor the performance and reliability of query responses, LlamaIndex's RetrieverEvaluator and custom metrics (i.e., Document Hit Rate, MMR) are used to assess accessibility and retrieval quality. The hybrid approach of retrieving and response evaluation is executed as shown in Figure \ref{fig:Evaluation Framework}.
\end{itemize}

\subsubsection{Document Ingestion and Indexing}
At the foundation of the LLM-Powered Adaptive Tutor is a document ingestion pipeline that parses domain-specific learning content, including structured PDFs and web-based curriculum material into chunked, semantically coherent segments. PDF parsing is handled using \texttt{PyPDF2}, a Python library designed for reading and extracting content from PDF files. It allows page-level control, making it suitable for extracting specific chapters or sections of interest. Alternative libraries such as \cite{yang2024extraction} (for layout-aware parsing) or \texttt{PyMuPDF} (for high-fidelity extraction) may also be considered for advanced use cases.

These coherent segments are then embedded into a vector database using OpenAI embeddings (\texttt{text-embedding-ada-002}). The LlamaIndex library manages this step, providing abstractions for:
\begin{itemize}
    \item Chunking strategies (fixed-length chunking is applied using SimpleNodeParser from LlamaIndex, which splits documents into chunks of 512 tokens)
    \item Embedding and metadata indexing, and
    \item Building a vector index that supports similarity search and metadata filtering.
\end{itemize}

\subsection{Experimental Results and Metrics }
In this personalized learning work, evaluation is performed on two complementary axes: the retriever's ability to surface relevant content and the LLM's ability to generate grounded, relevant responses. These are respectively handled by the Retriever Evaluator and the LLM Answering Evaluator.

As illustrated in Figure \ref{fig:Evaluation Framework}, the framework performs evaluation through two parallel pathways, synthetic QA-based evaluation, and manual query testing to assess both the retrieval and generation quality. This hybrid evaluation setup ensures that the system is rigorously tested for accuracy, context alignment, and robustness across multiple query types.

The Retriever Evaluator plays a critical role in assessing how effectively the system surfaces relevant document chunks in response to learner queries \cite{liu2009learning}. This evaluation is conducted through structured synthetic data, where QA pairs are automatically generated using document-ingested content with the help of LLM. These QA pairs act as ground truth to test whether the query engine can retrieve the exact source fragment that was used to formulate the answer.
This evaluation is independent of any language model output. The evaluation does not consider the final LLM answer; instead, it focuses solely on whether the retriever returns the correct supporting chunk that was originally used to generate the QA pair. By doing so, it isolates and quantifies the accuracy of the vector-based semantic search provided by LlamaIndex. Once queries are sent through, the retrieved chunks are compared with the known ground-truth chunk that originally came from the QA pair and database. This process enables a scalable and automated evaluation of retrieval performance.

To quantify how well the retriever performs, two metrics are applied: Hit Rate and Mean Reciprocal Rank (MRR).

\begin{table*}[]
\caption{Generated Queries set from Operating system data}
\renewcommand{\arraystretch}{1.2}
\resizebox{\textwidth}{!}{%
\begin{tabular}{cl}
\hline
\textbf{Queries} &
  \textbf{Description} \\ \hline
$Q_1$ &
  \begin{tabular}[c]{@{}l@{}}Discuss the challenges and setbacks faced during the development of the Multics project, and how these impacted its commercial success. \\ Include in your answer the role of different organizations such as IBM, General Electric, and Bell Labs.\end{tabular} \\ \hline
$Q_2$ &
  \begin{tabular}[c]{@{}l@{}}Explain the significance of the Multics project in the development of modern operating systems and security features. \\ Provide examples of its influence on subsequent systems, such as the UNIX system.\end{tabular} \\ \hline
$Q_3$ &
  \begin{tabular}[c]{@{}l@{}}Discuss the unique aspects of the Multics project in terms of its breadth of tasks, diversity of partners, and duration under development. \\ How did these factors contribute to the project's ability to pursue ambitious, long-term goals?\end{tabular} \\ \hline
$Q_4$ &
  \begin{tabular}[c]{@{}l@{}}Explain the process and the role of the supervisor when a user logs into a Multics system. \\ What are the steps involved in user authentication and how does the supervisor contribute to this process?\end{tabular} \\ \hline
$Q_5$ &
  \begin{tabular}[c]{@{}l@{}}Explain the fundamental concepts of the Multics system architecture, specifically focusing on the roles of processes and segments. \\ How do these elements interact within the system?\end{tabular} \\ \hline
$Q_6$ &
  \begin{tabular}[c]{@{}l@{}}What are protection rings in the context of the Multics system? \\ Discuss their hierarchical structure and their role in isolating the supervisor from other processes.\end{tabular} \\ \hline
$Q_7$ &
  \begin{tabular}[c]{@{}l@{}}Describe the actions that occur when a user logs into a Multics system. \\ What is the role of the trusted computing base (TCB) and the answering service in user authentication?\end{tabular} \\ \hline
$Q_8$ &
  \begin{tabular}[c]{@{}l@{}}Explain the process and significance of a Multics process requesting a segment that is not already in its descriptor segment. \\ What is the role of the descriptor segment and segment descriptor words (SDWs) in this process?\end{tabular} \\ \hline
$Q_9$ &
  \begin{tabular}[c]{@{}l@{}}Explain the fundamental concepts of the Multics system, specifically the roles of processes and segments. \\ How does a process's protection domain define the segments it can access?\end{tabular} \\ \hline
$Q_{10}$ &
  \begin{tabular}[c]{@{}l@{}}Describe the process by which a Multics process requests a segment that is not already in its descriptor segment. \\ Use the example of the segment named /U2/War/New Years Day to illustrate your answer.\end{tabular} \\ \hline
$Q_{11}$ &
  \begin{tabular}[c]{@{}l@{}}In the development and implementation of segmentation and classification models within image processing systems, how do users contribute to the accuracy of these processes? \\ Discuss the significance of user feedback and domain knowledge in enhancing model performance.\end{tabular} \\ \hline
$Q_{12}$ &
  \begin{tabular}[c]{@{}l@{}}Although 1x1 convolutions have proven valuable in reducing dimensionality and enhancing non-linearity in convolutional neural networks, \\ what are some potential shortcomings or limitations of using them in image processing tasks?\end{tabular} \\ \hline
$Q_{13}$ &
  What impact does the migration pattern of monarch butterflies have on the scheduling of machine learning model training cycles in distributed cloud environments? \\ \hline
$Q_{14}$ &
  How can large language models be leveraged to enhance cybersecurity education, detection, and threat analysis within modern digital systems? \\ \hline
$Q_{15}$ &
  \begin{tabular}[c]{@{}l@{}}What are the various categories of tools used in digital image processing, and how do they differ in terms of functionality and application? \\ Provide a detailed overview of key tools such as filtering techniques, morphological operations, and edge detection algorithms.\end{tabular} \\ \hline
\end{tabular}%
}
\label{tab:Queries Set}
\end{table*}

\begin{itemize}
    \item Hit Rate: measures whether the correct document chunk appears anywhere in the top-k retrieved results. It calculates the fraction of queries where the correct answer is found within the top-k retrieved documents. In simpler terms, it’s about how often our system gets it right within the top few guesses.
    
    In this work, if the chunk used to create the QA pair is found within the top-k results returned by the query engine, it is counted as a "hit", and the top-k setup is 5.
\[
\text{Hit Rate@}k = \frac{\text{Number of successful retrievals}}{\text{Total number of queries}}
\]

A high hit rate means the system can consistently retrieve supporting context from the vector database, which is essential for generating grounded and accurate LLM responses.

\item Mean Reciprocal Rank (MRR): MRR evaluates how highly ranked the first correct chunk is among the retrieved results. It serves as a metric for evaluating the accuracy of a system by examining the rank of the highest-placed relevant document for each query. Calculate the average of the reciprocals of these ranks in all queries. For example, if the first relevant document is ranked highest, the reciprocal rank is 1; if it is second, the reciprocal rank is 1/2, and so forth.

In this work: It checks the rank position at which the correct chunk appears for each query and computes the average reciprocal of those ranks.

\[
\text{MRR} = \frac{1}{N} \sum_{i=1}^{N} \frac{1}{\text{rank}_i}
\]

A higher MRR indicates that relevant documents are not just found but found early in the list, increasing their likelihood of influencing the final answer.

 To give an overview of how evaluation queries are selected. In our experiment, as shown in Figure \ref{fig:Evaluation Framework}, we generated a synthetic dataset by prompting the LLM to create two queries per chunk from the indexed document database. This process produced a diverse set of QA pairs to assess retrieval performance. Table \ref{tab:Queries Set} displays ten representative queries from this dataset, which are later used in Table \ref{tab:Retrieval Evaluation} to evaluate retrieval metrics. For LLM response evaluation shown in Table \ref{tab:Answer_eval}, the same queries were reused with the addition of five noise-injected queries to test the model’s robustness and response reliability.

Table \ref{tab:Queries Set} illustrates a few queries from the QA data pair set, showcasing a mix of queries grounded directly in the document content and those with slight variations in phrasing or focus. Once generated, all n queries were passed through the LlamaIndex query engine, which performed top-k retrieval based on semantic similarity. The system’s ability to locate the correct supporting chunk was then evaluated using Hit Rate and Mean Reciprocal Rank (MRR). Table \ref{tab:Retrieval Evaluation}, shows a sample of the same 10 evaluated queries, they are presented alongside their respective \texttt{MRR scores} and \texttt{Hit Rate scores}, highlighting how effectively the retriever surfaces relevant content under structured testing conditions.

The variation in MRR scores reflects the rank position at which the correct document chunk appears in the retrieval results. A perfect score of 1.0 indicates that the correct chunk was ranked first, while lower scores (e.g., 0.2, 0.5) mean the correct chunk was retrieved, but appeared lower in the top-k list. Queries with more ambiguous wording or overlap with multiple chunks tend to result in lower MRR due to reduced ranking confidence.
\end{itemize}

\begin{table}[]
\centering
\caption{Evaluation of Retrieval Performance Using Generated Queries}
\begin{tabular}{cccc}
\hline
Index & MRR      & \begin{tabular}[c]{@{}c@{}}Hit/Miss\end{tabular} & Questions \\ \hline
1     & 1.000000 & 1.0                                                 & $Q_1$     \\
2     & 0.200000 & 1.0                                                 & $Q_2$     \\
3     & 1.000000 & 1.0                                                 & $Q_3$     \\
4     & 0.500000 & 1.0                                                 & $Q_4$     \\
5     & 0.000000 & 0.0                                                 & $Q_5$     \\
6     & 1.000000 & 1.0                                                 & $Q_6$     \\
7     & 0.333333 & 1.0                                                 & $Q_7$    \\
8     & 1.000000 & 1.0                                                 & $Q_8$    \\
9     & 1.000000 & 1.0                                                 & $Q_9$    \\
10    & 0.200000 & 1.0                                                 & $Q_{10}$    \\ \hline
\end{tabular}
\label{tab:Retrieval Evaluation}
\end{table}

\begin{table*}[]
\caption{LLMs Response Evaluation on Generated queries}
\centering
\resizebox{\textwidth}{!}{%
\begin{tabular}{ccccccccccccc}
\hline
Models   & \multicolumn{3}{c}{GPT-3.5 Turbo} & \multicolumn{3}{c}{GPT-3.5 Turbo 16k} & \multicolumn{3}{c}{GPT-4} & \multicolumn{3}{c}{GPT-4 Turbo} \\ \hline
Response &
  \begin{tabular}[c]{@{}c@{}}Is \\ Irrelevant\end{tabular} &
  \begin{tabular}[c]{@{}c@{}}Relevancy \\ Score\end{tabular} &
  \begin{tabular}[c]{@{}c@{}}Faithfulness\\ Score\end{tabular} &
  \begin{tabular}[c]{@{}c@{}}Is \\ Irrelevant\end{tabular} &
  \begin{tabular}[c]{@{}c@{}}Relevancy \\ Score\end{tabular} &
  \begin{tabular}[c]{@{}c@{}}Faithfulness\\ Score\end{tabular} &
  Is Irrelevant &
  \begin{tabular}[c]{@{}c@{}}Relevancy \\ Score\end{tabular} &
  \begin{tabular}[c]{@{}c@{}}Faithfulness\\ Score\end{tabular} &
  \begin{tabular}[c]{@{}c@{}}Is \\ Irrelevant\end{tabular} &
  \begin{tabular}[c]{@{}c@{}}Relevancy \\ Score\end{tabular} &
  \begin{tabular}[c]{@{}c@{}}Faithfulness\\ Score\end{tabular} \\ \hline
$R_1$    & False       & 1.0      & 1.0      & False        & 1.0        & 1.0       & False    & 1.0    & 1.0   & False      & 1.0      & 1.0     \\
$R_1$    & False       & 1.0      & 1.0      & False        & 1.0        & 1.0       & False    & 1.0    & 1.0   & False      & 1.0      & 1.0     \\
$R_3$    & False       & 1.0      & 1.0      & False        & 1.0        & 1.0       & False    & 1.0    & 1.0   & False      & 1.0      & 1.0     \\
$R_4$    & False       & 1.0      & 1.0      & False        & 1.0        & 1.0       & False    & 1.0    & 1.0   & False      & 1.0      & 1.0     \\
$R_5$    & False       & 1.0      & 1.0      & False        & 1.0        & 1.0       & False    & 1.0    & 1.0   & False      & 1.0      & 1.0     \\
$R_6$    & False       & 1.0      & 1.0      & False        & 1.0        & 1.0       & False    & 1.0    & 1.0   & False      & 1.0      & 1.0     \\
$R_7$    & False       & 1.0      & 1.0      & False        & 1.0        & 1.0       & False    & 1.0    & 1.0   & False      & 1.0      & 1.0     \\
$R_8$    & False       & 1.0      & 1.0      & False        & 1.0        & 1.0       & False    & 1.0    & 1.0   & False      & 1.0      & 1.0     \\
$R_9$    & False       & 1.0      & 1.0      & False        & 1.0        & 1.0       & False    & 1.0    & 1.0   & False      & 1.0      & 1.0     \\
$R_{10}$    & False       & 1.0      & 1.0      & False        & 1.0        & 1.0       & False    & 1.0    & 1.0   & False      & 1.0      & 1.0     \\
$R_{11}$ & True        & 0.0      & 0.0      & True         & 0.0        & 0.0       & True     & 0.0    & 0.0   & True       & 0.0      & 0.0     \\
$R_{12}$ & True        & 0.0      & 0.0      & True         & 0.0        & 0.0       & True     & 0.0    & 0.0   & True       & 0.0      & 0.0     \\
$R_{13}$ & True        & 0.0      & 0.0      & True         & 0.0        & 0.0       & True     & 0.0    & 0.0   & True       & 0.0      & 0.0     \\
$R_{14}$ & True        & 0.0      & 0.0      & True         & 0.0        & 0.0       & True     & 0.0    & 0.0   & True       & 0.0      & 0.0     \\
$R_{15}$ & True        & 0.0      & 0.0      & True         & 0.0        & 0.0       & True     & 0.0    & 0.0   & True       & 0.0      & 0.0     \\ \hline
\end{tabular}%
}
\label{tab:Answer_eval}
\end{table*}

\subsubsection{LLM Answering Evaluation}
As illustrated in Figure \ref{fig:Evaluation Framework} In the LLM answering section, LLMs play a central role in interpreting queries and generating responses based on the retrieved contextual information. Once the top-k document chunks are retrieved by the system, the queries which are in the Table \ref{tab:Queries Set} , the same 10 queries used for Table \ref{tab:Retrieval Evaluation} retrieval evaluation along with some noise(i.e., irrelevant queries) and context are sent to one of the selected LLMs \texttt{GPT-3.5 Turbo}, \texttt{GPT-3.5 16k}, \texttt{GPT-4}, or \texttt{GPT-4 Turbo} to generate a final response. All 15 queries are sent through each model for evaluation and difference in response. 
They are evaluated on two critical axes: faithfulness and relevancy.

\begin{itemize}
    \item Faithfulness: Faithfulness measures whether the generated answers are grounded in the retrieved context. In other words, it measures whether the model’s answer is grounded in evidence and not hallucinated \cite{wu2024synchronous}. This is particularly critical in educational or high-stakes domains, where factual correctness is non-negotiable.

    \item Relevancy: Answer relevancy, on the other hand, measures how well the generated response addresses the user’s query. A response might be factually correct and even grounded in context, but if it doesn’t directly answer the question asked, its relevancy is low.
    This is especially useful when dealing with some noisy queries. For instance, if a question contains irrelevant terms or is poorly formed, a good LLM should still identify the intent and respond precisely.
\end{itemize}

Table \ref{tab:Answer_eval} displays responses for the questions from Table \ref{tab:Queries Set} stating relevancy and faithfulness score based on whether the queries are relevant or not.

A high faithfulness score of 1.0 means that the model did not invent or misinterpret facts, and strictly used the contextual information provided to form its answer. Faithfulness ensures accuracy and trust in LLM output.
A low relevancy can indicate that the model either: misunderstood the question, provided a vague or overly generic answer, or introduced unrelated information. So relevancy ensures usefulness and user alignment.

\begin{figure}[!h]
  \centering    \includegraphics[width=1\linewidth]{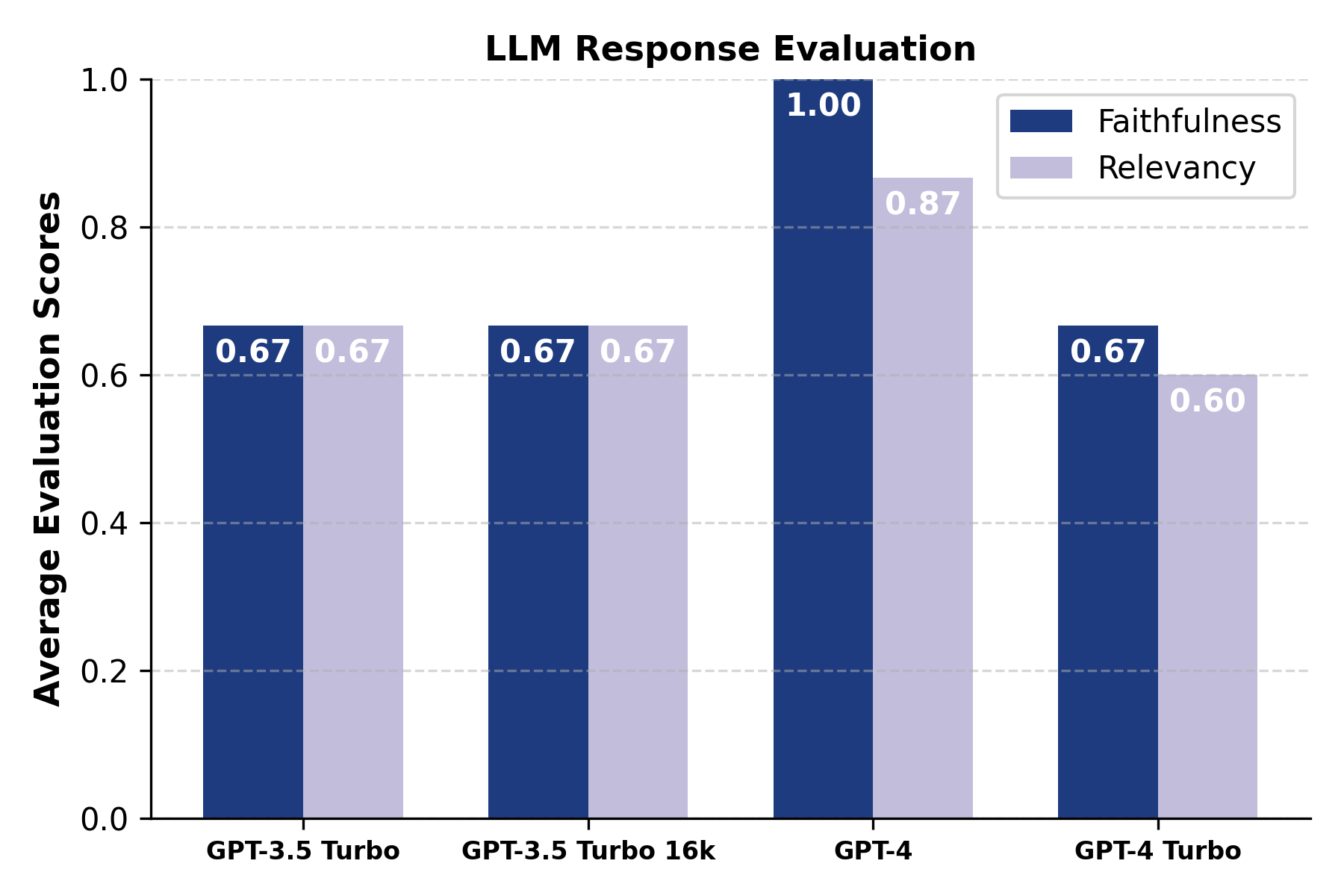}
    \caption{LLMs Response Evaluation based on Faithfulness and Relevancy}
    \label{fig:response_eval}
\end{figure}

The quantitative outputs of this LLM response evaluation are captured in the two plots:

The first plot Figure \ref{fig:response_eval} and Table \ref{tab:LLM_Eval average} displays the average LLM response evaluation score, where (GPT-4)stood out with a perfect faithfulness score of 1.0 and a high relevancy score of 0.87, indicating it consistently generated responses that were both accurate and well-aligned with retrieved context. \texttt{GPT-3.5} variants maintained consistent performance (both scoring 0.67), showing moderate alignment with expectations. Notably, \texttt{GPT-4 Turbo}, despite producing the longest responses, showed a drop in relevancy 0.60, suggesting potential issues with drifting from the query intent or including less focused information.

\begin{table}[]
\caption{Average LLM Response Evaluation Score}
\centering
\begin{tabular}{lll}
\hline
Model             & Relevancy Score & Faithfulness Score \\ \hline
GPT-3.5 Turbo     & 0.666667        & 0.666667           \\
GPT-3.5 Turbo 16k & 0.666667        & 0.666667           \\
GPT-4             & 0.933333        & 1.000000           \\
GPT-4 Turbo       & 0.666667        & 0.733333           \\ \hline
\end{tabular}
\label{tab:LLM_Eval average}
\end{table}

\begin{figure}[!h]
  \centering    \includegraphics[width=1\linewidth]{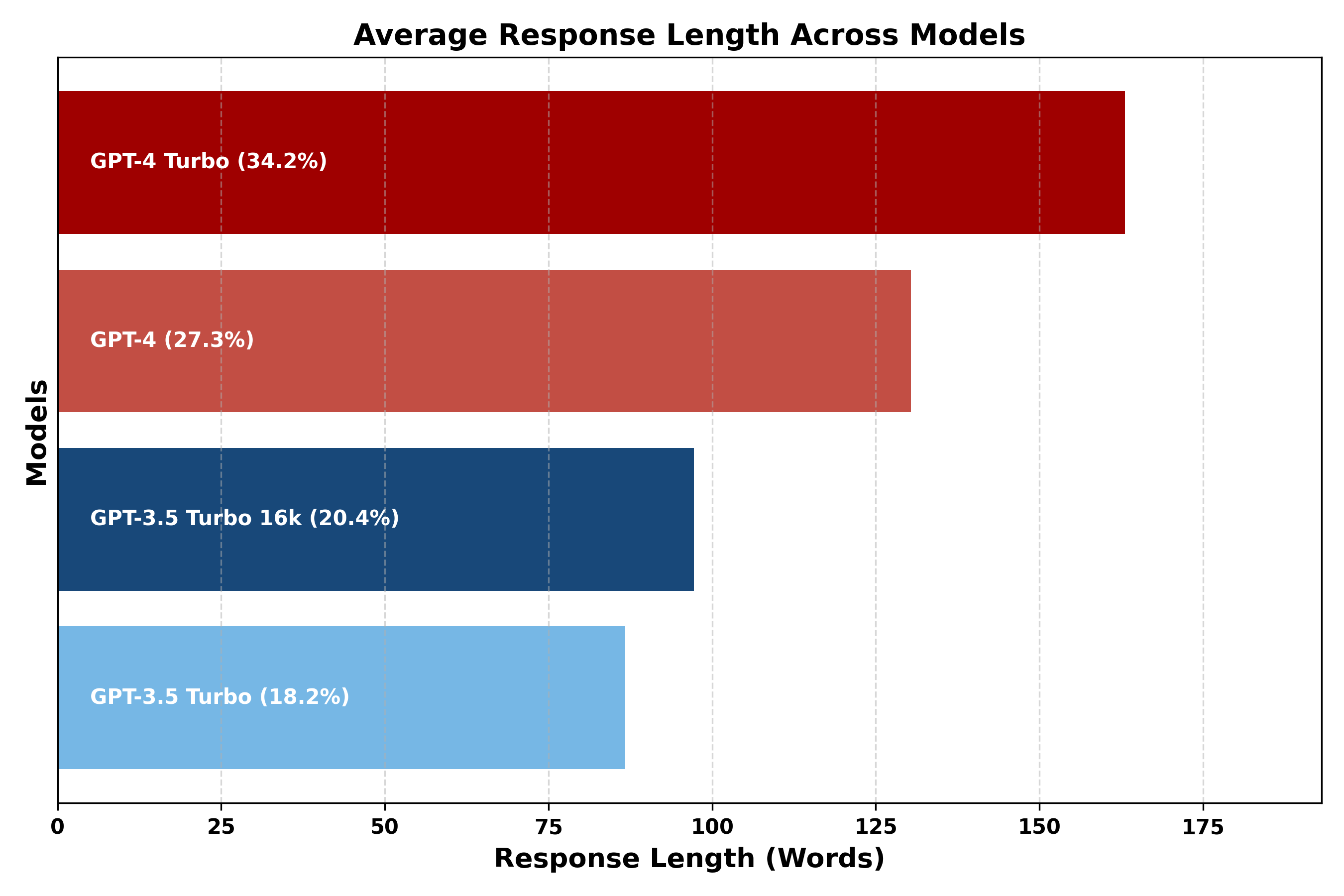}
    \caption{Response length across different GPT models}
    \label{fig:words}
\end{figure}

The second plot, Figure \ref{fig:words} displays the average response lengths generated by each model. Interestingly, \texttt{GPT-4 Turbo} produced the longest responses, followed by \texttt{GPT-3.5 Turbo 16k} and \texttt{GPT-4}, while \texttt{GPT-3.5 Turbo} responses were relatively concise. This may reflect differences in verbosity, token limit utilization, or inherent model tendencies in explanation depth. 

 These evaluations help determine not just which models are the most verbose, but which are the most dependable in retrieval-based QA settings, especially in personalized learning environments where factual precision and learner alignment are crucial.

\section{Conclusion and Future work} \label{sec:Conclusion}
This work highlights the importance of developing trustworthy, context-aware adaptive tutoring systems to support personalized learning at scale. By integrating vector-based retrieval with LLMs and employing evaluation metrics such as faithfulness and relevancy, the proposed RAG-PRISM framework ensures that learners receive accurate, grounded and contextually aligned feedback. This approach not only reduces misinformation, but also enables adaptive learner-centric instruction, bridging the gap between static digital platforms and intelligent, responsive learning environments.

The evaluation results demonstrated a well performance across both the retrieval and generation components. In our test case, the framework achieved a perfect document hit rate and Mean Reciprocal Rank (MRR) scores of 1.00, while GPT-4 outperformed other models with 100\% faithfulness and 93.3 \% relevancy. These results confirm the system’s capability to deliver high-quality, personalized cybersecurity instruction aligned with the learner context.

Future work will expand the applicability of the framework to real-world educational settings by incorporating inquiries from actual students and measuring their learning outcomes. In addition, we plan to diversify the content domains, explore alternative RAG implementations, and integrate human-in-the-loop feedback mechanisms. Scaling the evaluation across larger query sets will further strengthen the generalizability and performance assessment of the framework.

\section{Acknowledgment}
This work was partially supported by the National Science Foundation (NSF) under research projects 2335046, the University of Arizona's Provost Investment Fund for AI Hardware Design League (AI-HDL), and the OpenAI Researcher Access Program 0000011862.

\bibliographystyle{IEEEtran} 
\bibliography{refs}

\end{document}